\documentclass[11pt]{article}

\usepackage[english]{babel}
\usepackage[T1]{fontenc}
\usepackage[utf8]{inputenc}

\usepackage{amssymb,amsfonts,amsmath,amsthm}
\usepackage{stmaryrd}
\usepackage{braket}
\usepackage{physics}
\usepackage{enumerate}
\usepackage{xcolor}
\usepackage[top=1in,bottom=1in,left=1in,right=1in]{geometry}

\usepackage[colorlinks=true,linkcolor=blue,citecolor=blue,urlcolor=blue]{hyperref}

\newtheorem{theorem}{Theorem}
\newtheorem{proposition}{Proposition}
\newtheorem{lemma}{Lemma}
\newtheorem{corollary}{Corollary}

\newtheorem{assumption}{Assumption}
\newtheorem*{theorem-informal}{Theorem}

\begin{document}
\date{}
\title{Quantitative propagation of chaos for Lindblad dynamics} 


\author{Nina H. Amini\thanks{Université Paris-Saclay, CNRS,  CentraleSupélec, L2S, France. \texttt{nina.amini@centralesupelec.fr}} \and Sofiane Chalal\thanks{Université Paris-Saclay, CentraleSupélec, L2S, France. \texttt{sofiane.chalal@centralesupelec.fr}}}

\maketitle

\begin{abstract}
We consider an open quantum system governed $N$-body Lindblad equation and study mean-field limits in this setting.  We prove that the $N$-particle dynamics converges, in the sense of quantum relative entropy, to the tensorized solution of the limiting nonlinear equation. More precisely, we establish explicit bounds of order $1/N$ on the relative entropy between the $N$-particle density operator and the corresponding product state, thereby providing a quantitative propagation of chaos.
\end{abstract}


\medskip


\medskip

\tableofcontents

\section{Introduction}
\label{sec:intro}

\subsection{Motivation and setting}
{
The question addressed in this paper belongs to the general program of
statistical mechanics~\cite{kac56,mischler13kac}: deriving an effective
description of a large microscopic system from its many-body dynamics.
More precisely, one seeks to understand the asymptotic behavior of a system
consisting of a large number \(N\) of identical interacting quantum subsystems
as \(N\to\infty\). This amounts to establishing a law of large numbers for
quantum particles that are not independent, but whose correlations are expected
to become negligible at fixed order in a suitable scaling limit.

In the mean-field regime, each pair interaction is weak, of order \(1/N\),
while the cumulative effect of all interactions remains of order one.
The resulting macroscopic description is then given by a nonlinear one-body
equation, and the corresponding asymptotic independence property is known as
propagation of chaos.

Mean-field limits have therefore played a central role in the mathematical
description of large complex systems and have generated a vast literature. They
appear in kinetic theory and plasma physics through Vlasov-type equations, in
stellar dynamics through self-consistent gravitational models, in two-dimensional
fluid mechanics through point-vortex systems, and in probability theory through
interacting diffusions and McKean--Vlasov equations
\cite{braun77vlasov,
sznitman91topics,fournier14propagation}. In the
quantum setting, mean-field models arise naturally in the study of effective
dynamics for large bosonic systems, such as Bose--Einstein condensates
\cite{lieb05mathematics,bloch08many}. Their rigorous justification is typically
formulated through the notion of quantum propagation of chaos, introduced by
Spohn~\cite{spohn80kinetic}. Since then, many works have established mean-field
limits for closed quantum systems~\cite{bardos00weak,erdHos10derivation,rodnianski09quantum,knowles10mean,pickl11simple,golse19empirical,porat24pickl}.
By contrast, the open-system setting remains considerably less understood.

The first framework for open quantum mean-field dynamics goes back to Alicki and Messer~\cite{alicki83nonlinear}, who introduced the notion of a nonlinear quantum dynamical semigroup. This approach is particularly appealing because it places open quantum systems in close analogy with classical kinetic theory, where effective nonlinear equations emerge from many-particle dynamics in suitable scaling limits.

Two mean-field scenarios naturally arise in the open-system setting. The first concerns a collection of non-interacting quantum subsystems coupled to a common bath, and has been studied for instance in~\cite{merkil12,carollo24applicability}. In the present work,  we focus instead on the second scenario, namely a many-body system with mean-field interactions subject to dissipation.

More precisely, we consider an $N$-particle open quantum system whose state $\bar{\boldsymbol\rho}_t^N$ evolves according to the Lindblad master equation
\begin{align*}
\frac{\mathrm{d}\bar{\boldsymbol\rho}_t^N}{\mathrm{d}t}
=
-\mathrm{i}
\biggl[
\sum_{l=1}^N \tilde{\mathbf{H}}_l
+
\frac{1}{N}
\sum_{1\leq l<l'\leq N}
\mathbf A_{ll'},
\bar{\boldsymbol\rho}_t^N
\biggr]
+
\sum_{l=1}^N
\left(
\mathbf{L}_l \bar{\boldsymbol{\rho}}_t^N \mathbf{L}_l^\dagger
-
\frac{1}{2}
\{
\mathbf{L}_l^\dagger \mathbf{L}_l,
\bar{\boldsymbol\rho}_t^N
\}
\right).
\end{align*}

In the mean-field limits the nonlinear one-body dynamics. Formally, the limiting state $m_t$ solves the nonlinear Lindblad equation (see \cite{alicki83nonlinear})
\begin{align*}
\frac{\mathrm{d}m_t}{\mathrm dt}
=
-\mathrm{i}[\tilde H,m_t]
-
\mathrm{i}\,\mathrm{tr}_2
\bigl(
[A_{12},m_t\otimes m_t]
\bigr)
+
Lm_tL^\dagger
-
\frac{1}{2}
\{
L^\dagger L,m_t
\}.
\end{align*}

The purpose of this paper is to justify rigorously this mean-field limit through the notion of propagation of chaos. More precisely, we prove that if the initial $N$-particle state is approximately factorized, then this structure is propagated by the dynamics:
\begin{align*}
\bar{\boldsymbol\rho}_0^N
\approx
m_0^{\otimes N}
\qquad\Longrightarrow\qquad
\bar{\boldsymbol\rho}_t^N
\approx
m_t^{\otimes N},
\qquad
N\to\infty.
\end{align*}

Recently, Kolokoltsov derived in~\cite{kolokoltsov21law,kolokoltsov22qmfg} a related mean-field equation arising from systems of observed interacting quantum particles. However, propagation of chaos in that framework was established only in the pure bosonic setting, namely under symmetry assumptions at the level of wavefunctions. Since Lindblad dynamics is dissipative and does not preserve purity, it becomes necessary to work directly at the level of density operators in order to treat general mixed states.

The main contribution of the present work is to establish quantitative propagation of chaos directly at the level of density matrices. More precisely, we obtain explicit convergence rates of order $1/N$ in normalized quantum relative entropy.
}

\subsection{Entropy method and main result}
\label{subsec:intro_method}

Our approach is inspired by the relative entropy method introduced by Jabin and Wang~\cite{jabin16mean,jabin18,saint19points} in the classical setting of mean-field limits for interacting diffusion systems. One of the strengths of this method is that it provides quantitative estimates naturally adapted to mean-field interactions.

In the classical framework, the $N$-particle density $\bar\nu_t^N$ is compared directly with the tensorized limit density $\bar\mu_t^{\otimes N}$ through the normalized relative entropy
\begin{align*}
\mathfrak{H}_N^{\mathrm{c}}(t) := \frac{1}{N}\mathfrak{D}^{\mathrm{c}}\bigl(\bar{\nu}_t^N \big\| \bar{\mu}_t^{\otimes N} \bigr),
\end{align*}
where $\mathfrak{D}^{\mathrm{c}}$ denotes the classical relative entropy. Under suitable assumptions, this method yields estimates of the form
\begin{align*}
\mathfrak{H}_N^{\mathrm{c}}(t)
\leq e^{Ct}
\left( \mathfrak{H}_N^{\mathrm{c}}(0) + \frac{1}{N} \right),
\end{align*}
which imply propagation of chaos through Pinsker's inequality.

The point of view of the present paper is to transpose this strategy to the quantum setting, replacing
\begin{align*}
\bar{\nu}_t^N
\leadsto
\bar{\boldsymbol\rho}_t^N,
\qquad
\bar{\mu}_t^{\otimes N}
\leadsto
m_t^{\otimes N},
\qquad
\mathfrak{D}^{\mathrm{c}}
\leadsto
\mathfrak{D},
\end{align*}
where $\mathfrak{D}$ denotes the Umegaki quantum relative entropy~\cite{umegaki62conditional}. This choice is dictated by two fundamental properties: monotonicity under completely positive trace-preserving maps and compatibility with the quantum Pinsker inequality.

The main difficulty in adapting the entropy method to the quantum framework lies in the noncommutative structure of the interaction terms. Our proof proceeds by differentiating the normalized relative entropy along the dynamics and exploiting the monotonicity of the Umegaki relative entropy under the Lindblad semigroup in order to absorb the dissipative contributions. The interaction terms are then reduced to centered two-body observables, whose contribution is estimated through exponential moment bounds combined with a combinatorial analysis of the surviving interaction configurations.

We now state informally the main result of the paper.

\begin{theorem-informal}
Assume that the initial non-linear state  faithful and the 
$N$-body state  exchangeable. Then the normalized quantum relative entropy
$$
\mathfrak H_N(t)
=
\frac{1}{N}
\mathfrak{D}
\bigl(
\bar{\boldsymbol{\rho}}_t^N
\,\big\|\,
m_t^{\otimes N}
\bigr)
$$
satisfies an estimate of the form
$$
\mathfrak H_N(t)
\leq
e^{Ct}
\left(
\mathfrak H_N(0)
+
\frac{C}{N}
\right),
\qquad
t\in[0,T],
$$
where $C>0$ depends only on $m_{0},\|L\|,\|A\|,$ and $T$.

As a consequence, for every fixed $k\geq1$,
$$
\sup_{t\in[0,T]}
\left\|
\bar{\boldsymbol\rho}_t^{N:(k)} - m_t^{\otimes k}
\right\|_1
\longrightarrow0,
\qquad
N\to\infty.
$$
In other words, the Lindblad dynamics exhibits quantum propagation of chaos uniformly on finite time intervals.
\end{theorem-informal}

The precise quantitative statement is given later in Theorem~\ref{thm:quantum_entropy_poc}.
To the best of our knowledge, this is the first quantitative propagation-of-chaos result for mean-field Lindblad dynamics obtained directly at the level of density operators.

\paragraph{Organization of the paper.}

The paper is organized as follows. Section~\ref{sec:notation} introduces the notation and preliminary material used throughout the paper. Section~\ref{sec:framework} presents the $N$-particle dynamics and the associated nonlinear mean-field equation. Section~\ref{sec:entropy_method} contains the statement and proof of the main entropy estimate. The appendices collect auxiliary identities together with the proofs of the structural and combinatorial lemmas.

\section{Preliminaries and notation}
\label{sec:notation}

We fix throughout the paper a finite time horizon $T>0$. The symbol $\mathrm{i}$ denotes the imaginary unit.

 We work throughout the paper on a finite-dimensional complex Hilbert space
\(\mathbb H\), endowed with its canonical Hermitian inner product
\(\langle\cdot,\cdot\rangle\) (linear in the second argument) and norm
\[
\|\psi\|=\sqrt{\langle\psi,\psi\rangle}.
\]
The space of bounded linear operators on \(\mathbb H\) is denoted by
\(\mathcal B(\mathbb H)\) and equipped with the operator norm
\begin{align*}
\|O\|
:=
\sup_{\|\psi\|\le 1}\|O\psi\|.
\end{align*}
When \(\dim\mathbb H=d\), we identify
\[
\mathcal B(\mathbb H)\simeq\mathcal M_d(\mathbb C).
\]
The identity operator is denoted by \(\mathbf 1\), and the trace of
\(O\in\mathcal B(\mathbb H)\) by \(\mathrm{tr}(O)\).
The adjoint of \(O\) is denoted by \(O^\dagger\), and \(O\) is said to be
\emph{self-adjoint} if \(O=O^\dagger\).

The trace norm is defined by
\begin{align*}
\|O\|_1
:=
\mathrm{tr}\bigl(\sqrt{O^\dagger O}\bigr).
\end{align*}

For a self-adjoint operator $O\in\mathcal{B}(\mathbb{H})$, we denote by 
$\lambda_{\min}(O)$ its smallest eigenvalue. The commutator and the anticommutator of $O_{1},O_{2}\in\mathcal{B}(\mathbb{H})$ are
\begin{align*}
[O_{1},O_{2}]:=O_{1}O_{2}-O_{2}O_{1},
\qquad
\{O_{1},O_{2}\}:=O_{1}O_{2}+O_{2}O_{1}.
\end{align*}
The set of \emph{quantum states}, or density operators, is
\begin{align*}
\mathcal{S}(\mathbb{H}):=\bigl\{\rho\in\mathcal{B}(\mathbb{H}):\rho=\rho^{\dagger},\;\rho\geq 0,\;\mathrm{tr}(\rho)=1\bigr\}.
\end{align*}
We say that a state \(\rho\in\mathcal S(\mathbb H)\) is
\emph{faithful} if \(\rho>0\). The support
\(\mathrm{supp}(\rho)\) of a state \(\rho\) is the orthogonal projection
onto the range of \(\rho\). Moreover, \(\rho\) is faithful if and only if
$
\mathrm{supp}(\rho)=\mathbf 1.
$

For an integer $N\geq 1$, the $N$-fold tensor product $\mathbb{H}^{\otimes N}$ is itself a Hilbert space, equipped with the canonical tensor inner product. For $O\in\mathcal{B}(\mathbb{H})$ and $l\in\{1,\dots,N\}$, we write
\begin{align*}
\mathbf{O}_{l}:=\mathbf{1}\otimes\cdots\otimes O\otimes\cdots\otimes\mathbf{1}\in\mathcal{B}(\mathbb{H}^{\otimes N})
\end{align*}
for the operator acting as $O$ on the $l$-th factor and as the identity on the others. Similarly, for $B\in\mathcal{B}(\mathbb{H}\otimes\mathbb{H})$ and $l\neq l'$, $\mathbf{B}_{ll'}\in\mathcal{B}(\mathbb{H}^{\otimes N})$ acts as $B$ on the $l$-th and $l'$-th factors. Given $\rho_{12}\in\mathcal{B}(\mathbb{H}\otimes\mathbb{H})$, the \emph{partial trace} over the second factor is the unique linear map $\mathrm{tr}_{2}:\mathcal{B}(\mathbb{H}\otimes\mathbb{H})\to\mathcal{B}(\mathbb{H})$ characterized by $\mathrm{tr}_{2}(A\otimes B)=\mathrm{tr}(B)A$. Partial traces over arbitrary subsets of factors of $\mathbb{H}^{\otimes N}$ are defined analogously.

A state $\bar{\boldsymbol\rho}^{N}\in\mathcal{S}(\mathbb{H}^{\otimes N})$ is \emph{exchangeable} if it is invariant under arbitrary permutations of the $N$ factors. The $k$-th marginal $\bar{\boldsymbol\rho}^{N:(k)}\in\mathcal{S}(\mathbb{H}^{\otimes k})$ is obtained by tracing out any $N-k$ factors; for an exchangeable state, this does not depend on the choice. More generally, given a subset $J\subseteq\{1,\dots,N\}$ of size $k$, we denote by $\bar{\boldsymbol\rho}^{N:(k),J}$ the marginal obtained by tracing out the factors not in $J$.

A linear map $\Phi:\mathcal{B}(\mathbb{H})\to\mathcal{B}(\mathbb{H})$ is said to be \emph{completely positive and trace-preserving} (CPTP) if $\Phi\otimes\mathbf{1}_{\mathcal{B}(\mathbb{H}')}$ is positive for every auxiliary Hilbert space $\mathbb{H}'$, and $\mathrm{tr}(\Phi(O))=\mathrm{tr}(O)$ for every $O\in\mathcal{B}(\mathbb{H})$. 

For $\sigma\in\mathcal{B}(\mathbb{H})$ self-adjoint with $\sigma>0$, $\log\sigma$ is 
defined via the functional calculus, and the Fréchet differential of the logarithm at 
$\sigma$ is the linear map $\mathcal{T}_{\sigma}:\mathcal{B}(\mathbb{H})\to\mathcal{B}(\mathbb{H})$ 
given by
\begin{align}\label{eq:def_T_sigma}
\mathcal{T}_{\sigma}(X):=\int_{0}^{\infty}(\sigma+s\mathbf{1})^{-1}X(\sigma+s\mathbf{1})^{-1}\,\mathrm{d}s.
\end{align}
For every $C^{1}$ curve $t\mapsto\sigma_{t}>0$,
\begin{align}\label{eq:dt_log}
\frac{\mathrm{d}}{\mathrm{d}t}\log\sigma_{t}=\mathcal{T}_{\sigma_{t}}(\dot\sigma_{t}),
\end{align}
and for every $X,Y\in\mathcal{B}(\mathbb{H})$,
\begin{align}
\mathcal{T}_{\sigma}\bigl([X,\sigma]\bigr)&=[X,\log\sigma],\label{eq:T_commutator}\\
\mathcal{T}_{\sigma}(\sigma)&=\mathbf{1},\label{eq:T_sigma_identity}\\
\mathrm{tr}\bigl(Y\,\mathcal{T}_{\sigma}(X)\bigr)&=\mathrm{tr}\bigl(\mathcal{T}_{\sigma}(Y)X\bigr).\label{eq:T_self_adjoint}
\end{align}
In particular,
\begin{align}\label{eq:trace_T}
\mathrm{tr}\bigl(\sigma\,\mathcal{T}_{\sigma}(X)\bigr)=\mathrm{tr}(X).
\end{align}

We recall the definition and the basic properties of the quantum relative entropy. For a comprehensive treatment, we refer to~\cite{ohya04quantum}.

For $\rho,\sigma\in\mathcal{S}(\mathbb{H})$, the \emph{quantum relative entropy} of $\rho$ with respect to $\sigma$ is
\begin{align}\label{eq:def_relative_entropy}
\mathfrak{D}(\rho\|\sigma):=
\begin{cases}
\mathrm{tr}\bigl(\rho(\log\rho-\log\sigma)\bigr) & \text{if }\mathrm{supp}(\rho)\subseteq\mathrm{supp}(\sigma),\\
+\infty & \text{otherwise.}
\end{cases}
\end{align}
When $\sigma$ is faithful, $\mathfrak{D}(\rho\|\sigma)$ is finite for every $\rho$. The relative entropy is non-negative and vanishes if and only if $\rho=\sigma$.

Two structural properties will be used repeatedly. The first is the \emph{monotonicity under CPTP maps}: for every CPTP map $\Phi$ and every $\rho,\sigma\in\mathcal{S}(\mathbb{H})$,
\begin{align}\label{eq:monotonicity}
\mathfrak{D}\bigl(\Phi(\rho)\|\Phi(\sigma)\bigr)\leq\mathfrak{D}(\rho\|\sigma).
\end{align}

A direct consequence, applied to the partial trace, yields the \emph{monotonicity under reduction}:
\begin{align}\label{eq:monotonicity_partial_trace}
\mathfrak{D}\bigl(\mathrm{tr}_{2}(\rho_{12})\,\big\|\,\mathrm{tr}_{2}(\sigma_{12})\bigr)\leq\mathfrak{D}(\rho_{12}\|\sigma_{12}).
\end{align}
The second is the \emph{superadditivity with respect to tensorized reference states}: for every $\rho^{N}\in\mathcal{S}(\mathbb{H}^{\otimes N})$, every $\sigma\in\mathcal{S}(\mathbb{H})$, and every decomposition of $\{1,\dots,N\}$ into $n$ disjoint blocks $B_{1},\dots,B_{n}$ of size $k$ (so that $N=nk$),
\begin{align}\label{eq:superadditivity}
\mathfrak{D}\bigl(\rho^{N}\,\big\|\,\sigma^{\otimes N}\bigr)\geq\sum_{j=1}^{n}\mathfrak{D}\bigl(\rho^{N:(k),B_{j}}\,\big\|\,\sigma^{\otimes k}\bigr),
\end{align}
where $\rho^{N:(k),B_{j}}$ is the marginal of $\rho^{N}$ on the block $B_{j}$.

We will also use a variational formula and the quantum Pinsker inequality. The \emph{variational formula}: for every $\rho\in\mathcal{S}(\mathbb{H})$, every faithful $\sigma\in\mathcal{S}(\mathbb{H})$, every self-adjoint $X\in\mathcal{B}(\mathbb{H})$, and every $\lambda>0$,
\begin{align}\label{eq:variational}
\mathrm{tr}(\rho X)\leq\frac{1}{\lambda}\,\mathfrak{D}(\rho\|\sigma)+\frac{1}{\lambda}\,\log\mathrm{tr}\bigl(e^{\log\sigma+\lambda X}\bigr).
\end{align}
The \emph{quantum Pinsker inequality}:
\begin{align}\label{eq:pinsker}
\frac{1}{2}\,\|\rho-\sigma\|_{1}^{2}\leq\mathfrak{D}(\rho\|\sigma).
\end{align}

We finally recall the \emph{Golden--Thompson inequality}: for every self-adjoint $A,B\in\mathcal{B}(\mathbb{H})$,
\begin{align}\label{eq:golden_thompson}
\mathrm{tr}\bigl(e^{A+B}\bigr)\leq\mathrm{tr}\bigl(e^{A}e^{B}\bigr).
\end{align}
Combining~\eqref{eq:variational} and~\eqref{eq:golden_thompson} with $A=\log\sigma$ and $B=\lambda X$,
\begin{align}\label{eq:variational_GT}
\mathrm{tr}(\rho X)\leq\frac{1}{\lambda}\,\mathfrak{D}(\rho\|\sigma)+\frac{1}{\lambda}\,\log\mathrm{tr}\bigl(\sigma\,e^{\lambda X}\bigr),
\end{align}
which is the form we shall apply in Theorem~\ref{thm:quantum_entropy_poc}.
\section{Model and mean-field limit}
\label{sec:framework}
\subsection{The $N$-particle Lindblad dynamics}
The model is a system of $N$ identical quantum particles, where each pair of particles interacts with equal strength, independently of their labels. The interaction strength is scaled by $1/N$ so that the total energy
remains of order $N$. The particles are labeled $1,\dots,N$, each carrying a copy $\mathbb{H}_l\simeq\mathbb{H}.$

The model is specified by three operators:
\begin{itemize}
    \item a self-adjoint one-body Hamiltonian $\tilde H\in\mathcal{B}(\mathbb{H})$;
    \item a self-adjoint two-body interaction
    $A\in\mathcal{B}(\mathbb{H}\otimes\mathbb{H})$, which we further assume
    to be {exchange-symmetric}\footnote{The exchange-symmetry assumption~\eqref{eq:swap_symmetry} reflects the
indistinguishability of the particles at the level of the interaction: two
particles interact in the same way whichever of them we label first. }, that is,
    \begin{align}\label{eq:swap_symmetry}
    \mathrm{S}\,A\,\mathrm{S}=A,
    \end{align}
    where $\mathrm{S}\in\mathcal{B}(\mathbb{H}\otimes\mathbb{H})$ denotes
    the flip operator $\psi\otimes\varphi\mapsto\varphi\otimes\psi$;
    \item a Lindblad jump operator $L\in\mathcal{B}(\mathbb{H})$, modeling
    the local coupling of each particle to its environment.
\end{itemize}

The total $N$-body Hamiltonian is
\begin{align}\label{eq:H_N}
\mathbf{H}^{N}
:=
\sum_{l=1}^{N}\tilde{\mathbf{H}}_{l}
+\frac{1}{N}\sum_{1\leq l<l'\leq N}\mathbf{A}_{ll'},
\end{align}
and the $N$-body open dynamics is governed by the Lindblad master equation
\begin{align}\label{eq:N_body_lindblad}
\frac{\mathrm{d}\bar{\boldsymbol\rho}_{t}^{N}}{\mathrm{d}t}
=
\mathcal{L}^{N}(\bar{\boldsymbol\rho}_{t}^{N})
:=
-\mathrm{i}[\mathbf{H}^{N},\bar{\boldsymbol\rho}_{t}^{N}]
+\sum_{l=1}^{N}\Bigl(
\mathbf{L}_{l}\,\bar{\boldsymbol\rho}_{t}^{N}\,\mathbf{L}_{l}^{\dagger}
-\tfrac{1}{2}\bigl\{\mathbf{L}_{l}^{\dagger}\mathbf{L}_{l},\bar{\boldsymbol\rho}_{t}^{N}\bigr\}
\Bigr).
\end{align}
The generator $\mathcal{L}^{N}$ is a bounded linear map on
$\mathcal{B}(\mathbb{H}^{\otimes N})$, and~\eqref{eq:N_body_lindblad} admits a
unique global solution in $\mathcal{S}(\mathbb{H}^{\otimes N})$ for every initial
datum. The associated semigroup $(e^{t\mathcal{L}^{N}})_{t\geq 0}$ is CPTP. Both
$\mathbf{H}^{N}$ and $\sum_{l}\mathcal{D}_{l}$ (where
$\mathcal{D}_{l}(\cdot):=\mathbf{L}_{l}\cdot\mathbf{L}_{l}^{\dagger}-\tfrac{1}{2}\{\mathbf{L}_{l}^{\dagger}\mathbf{L}_{l},\cdot\}$
denotes the local dissipator at site $l$) are permutation-invariant, so
exchangeability is preserved by the flow.

\begin{assumption}[Initial chaos]\label{ass:initial-chaos}
There exists $\varrho_{0}\in\mathcal{S}(\mathbb{H})$ such that
\begin{align*}
\bar{\boldsymbol\rho}_{0}^{N}=\varrho_{0}^{\otimes N}.
\end{align*}
\end{assumption}

In particular, $\bar{\boldsymbol\rho}_{0}^{N}$ is exchangeable. Since $\mathcal{L}^{N}$ is permutation-invariant, it follows that $\bar{\boldsymbol\rho}_{t}^{N}$ remains exchangeable for every $t\in[0,T]$.

\subsection{Mean-field limit equation}
\label{subsec:mean_field_dynamics}

We now describe the limiting nonlinear dynamics expected at large $N$. We first
introduce the mean-field self-interaction associated with $A$. 

\paragraph{The mean-field self-interaction.}
Given $\sigma\in\mathcal{S}(\mathbb{H})$, we define
\begin{align}\label{eq:def_A_sigma}
A^{\sigma}:=\mathrm{tr}_{2}\bigl((\mathbf{1}\otimes\sigma)A\bigr)\in\mathcal{B}(\mathbb{H}).
\end{align}
The map $\sigma\mapsto A^{\sigma}$ is linear, $A^{\sigma}$ is self-adjoint, and
$\|A^{\sigma}\|\leq\|A\|$. The operator $A^{\sigma}$ encodes the
\emph{average potential} felt by a single particle in the state $\sigma$ when
it interacts via $A$ with another particle independently distributed
according to $\sigma$.

The limiting nonlinear dynamics is given by the mean-field Lindblad equation
\begin{align}\label{eq:mf_lindblad}
\frac{\mathrm{d}m_{t}}{\mathrm{d}t}
=
-\mathrm{i}[\tilde H + A^{m_t},m_{t}]
+L\,m_{t}\,L^{\dagger}
-\tfrac{1}{2}\{L^{\dagger}L,m_{t}\}.
\end{align}
Under the exchange-symmetry condition~\eqref{eq:swap_symmetry}, the interaction term
can be rewritten in terms of a two-body commutator. More precisely, for every 
$\sigma\in\mathcal{S}(\mathbb{H})$,
\begin{align}\label{eq:mf_identity}
\mathrm{tr}_{2}\bigl([A_{12},\sigma\otimes\sigma]\bigr)=[A^{\sigma},\sigma].
\end{align}
As a consequence, equation~\eqref{eq:mf_lindblad} can equivalently be written as
\begin{align}\label{eq:mf_lindblad_partial_trace}
\frac{\mathrm{d}m_{t}}{\mathrm{d}t}
=
-\mathrm{i}[\tilde{H},m_{t}]
-\mathrm{i}\,\mathrm{tr}_{2}\bigl([A_{12},m_{t}\otimes m_{t}]\bigr)
+L\,m_{t}\,L^{\dagger}
-\tfrac{1}{2}\{L^{\dagger}L,m_{t}\}.
\end{align}

\subsection{Logarithmic bounds}

In the classical relative entropy method, a common assumption to obtain an explicit estimate is that the limiting
density is bounded away from zero (see~\cite[Theorem 1 \& Remark 2]{jabin18}). This ensures that its logarithm is well-defined
and uniformly controlled. In the finite-dimensional quantum setting, the analogous
condition is the faithfulness of the limiting one-body state $m_t$. In the following, we only assume that the initial one-body state $m_0$ is faithful; Proposition~\ref{prop:faithfulness} below shows that this property is then automatically propagated by the mean-field flow.

\begin{assumption}[Faithful initial state]
\label{ass:faithful}
The initial one-body state $m_{0}\in\mathcal{S}(\mathbb{H})$ is faithful, namely
$$
m_0>0.
$$
\end{assumption}

\begin{proposition}
\label{prop:faithfulness}
Let $(m_{t})_{t\in[0,T]}$ be the solution to~\eqref{eq:mf_lindblad}. Then $m_{t}$ is faithful for every $t\in[0,T]$, and
\begin{align}\label{eq:lower_bound_eigenvalue}
\lambda_{\min}(m_{t})
\geq
\lambda_{\min}(m_{0})\,e^{-\|L\|^{2}t}.
\end{align}
In particular, $\log m_t$ is well-defined for all $t\in[0,T]$, and
\begin{align}\label{eq:bound_Lambda_T}
\sup_{0\leq t\leq T}\|\log m_{t}\|
\leq
-\log\lambda_{\min}(m_{0})
+
\|L\|^{2}T.
\end{align}
\end{proposition}

The proof, given in Appendix~\ref{app:proof_faithfulness}, relies on a Duhamel representation of the mean-field flow, expressing $m_{t}$ as the sum of a propagated initial term and a non-negative dissipative contribution.

\section{Quantum propagation of chaos via relative entropy}
\label{sec:entropy_method}

\subsection{Main statements}
\label{subsec:statements}

Throughout this section we assume Assumptions~\ref{ass:initial-chaos} 
and~\ref{ass:faithful}. In particular, by Proposition~\ref{prop:faithfulness}, 
the solution $(m_t)_{t\in[0,T]}$ remains faithful, so that 
$\Lambda_t := \log m_t$ is well-defined for all $t\in[0,T]$.

We measure the deviation between the $N$-particle dynamics and the mean-field 
limit using the normalized relative entropy
\begin{align}\label{eq:def_HN}
\mathfrak{H}_{N}(t):=\frac{1}{N}\,\mathfrak{D}\bigl(\bar{\boldsymbol\rho}^{N}_t\,\big\|\,m^{\otimes N}_t\bigr).
\end{align}

The following theorem provides a quantitative control of $\mathfrak{H}_N(t)$.

For convenience, we set
\begin{align}\label{eq:def_K}
\mathrm{K}:=-\log\lambda_{\min}(m_{0})+\|L\|^{2}T,
\end{align}
which by Proposition~\ref{prop:faithfulness} provides an upper bound for $\sup_{t\in[0,T]}\|\Lambda_{t}\|$.

\begin{theorem}[Quantitative entropy estimate]\label{thm:quantum_entropy_poc}
Under Assumptions~\ref{ass:initial-chaos} and~\ref{ass:faithful}, for every 
$0<\mathfrak{q}<\frac{e^{-2}}{8\|A\|\mathrm{K}}$, every $t\in[0,T]$, and every $N\geq 1$,
\begin{align}\label{eq:main_estimate}
\mathfrak{H}_{N}(t)\leq e^{t/\mathfrak{q}}\!\left(\mathfrak{H}_{N}(0)+\frac{\log C_{T,\mathfrak{q}}+2\mathfrak{q}\|A\|\mathrm{K}}{N}\right),
\end{align}
where
\begin{align}\label{eq:def_C_T_q}
C_{T,\mathfrak{q}}:=1+\frac{x_{T,\mathfrak{q}}}{(1-x_{T,\mathfrak{q}})^{2}}+\frac{y_{T,\mathfrak{q}}}{1-y_{T,\mathfrak{q}}},\quad \textrm{with}\quad x_{T,\mathfrak{q}}:=8e^{2}\mathfrak{q}\|A\|\mathrm{K}\quad \textrm{and}\quad y_{T,\mathfrak{q}}:=16e\mathfrak{q}\|A\|\mathrm{K}.
\end{align}
\end{theorem}

When the initial datum is exactly chaotic, $\bar{\boldsymbol\rho}_{0}^{N}=m_{0}^{\otimes N}$, 
one has $\mathfrak{H}_{N}(0)=0$, and the estimate yields a $\mathcal{O}(1/N)$ bound, 
uniformly on $[0,T]$.

\begin{corollary}[Propagation of chaos]\label{cor:entropy_to_marginals}
Under the assumptions of Theorem~\ref{thm:quantum_entropy_poc}, assume in addition that 
$\mathfrak{H}_{N}(0)\to 0$ as $N\to\infty$. Then, for every fixed $k\geq 1$,
\begin{align}\label{eq:cor_entropy_marginal}
\sup_{t\in[0,T]}\mathfrak{D}\bigl(\bar{\boldsymbol\rho}_{t}^{N:(k)}\,\big\|\,m_{t}^{\otimes k}\bigr)\xrightarrow[N\to\infty]{}0,
\end{align}
and consequently
\begin{align}\label{eq:cor_trace_norm}
\sup_{t\in[0,T]}\bigl\|\bar{\boldsymbol\rho}_{t}^{N:(k)}-m_{t}^{\otimes k}\bigr\|_{1}\xrightarrow[N\to\infty]{}0.
\end{align}
In particular, $(\bar{\boldsymbol\rho}_{t}^{N})_{N\geq 1}$ is $m_{t}$-chaotic, 
uniformly on $[0,T]$.
\end{corollary}

\subsection{Preliminary lemmas}
\label{subsec:combinatorial_lemmas}

Before turning to the proof of Theorem~\ref{thm:quantum_entropy_poc}, we isolate two 
combinatorial lemmas that lie at the heart of the argument.

\begin{lemma}\label{lem:isolated_vertex}
Let $h\in\mathcal{B}(\mathbb{H}\otimes\mathbb{H})$ and 
$\varrho\in\mathcal{S}(\mathbb{H})$ satisfy the following assumption
\begin{align}\label{eq:centering_assumption}
\mathrm{tr}_{2}\bigl((\mathbf{1}\otimes\varrho)h\bigr)=0,
\qquad
\mathrm{tr}_{1}\bigl((\varrho\otimes\mathbf{1})h\bigr)=0.
\end{align}
For each edge $\mathrm{e}=\{i,j\}$ with $1\leq i<j\leq N$, denote by $h_{\mathrm{e}}$ the copy of $h$ 
acting on the $i$-th and $j$-th tensor factors of $\mathbb{H}^{\otimes N}$. 

Then, for every product $h_{\mathrm{e}_{1}}\cdots h_{\mathrm{e}_{k}}$, if there exists an index 
$a\in\{1,\dots,N\}$ that appears as an endpoint of exactly one of the edges 
$\mathrm{e}_{1},\dots,\mathrm{e}_{k}$, one has
\begin{align}\label{eq:isolated_vanishing}
\mathrm{tr}\bigl(\varrho^{\otimes N}\,h_{\mathrm{e}_{1}}\cdots h_{\mathrm{e}_{k}}\bigr)=0.
\end{align}
\end{lemma}

The proof of 
Lemma~\ref{lem:isolated_vertex}, obtained by isolating the unique factor acting 
non-trivially on the marked site, is given in Appendix~\ref{app:proof_isolated_vertex}.

The next lemma quantifies how rare the surviving configurations are.

\begin{lemma}\label{lem:endpoint_counting}
For $N,k\geq 1$, define
\begin{align}\label{eq:def_K_set}
\mathfrak{K}_{N,2k}:=\Bigl\{
(i_{1},j_{1},\dots,i_{k},j_{k})\in\{1,\dots,N\}^{2k}
:\text{no index appears exactly once}
\Bigr\}.
\end{align}
Then:
\begin{enumerate}
\item If $2k\leq N$, then
\[
|\mathfrak{K}_{N,2k}|\leq k\,e^{k}\,N^{k}\,k^{k}.
\]
\item If $2k>N$, then
\[
|\mathfrak{K}_{N,2k}|\leq N^{2k}.
\]
\end{enumerate}
\end{lemma}

The proof, in the regime $2k\leq N$, decomposes the configurations according to the 
number $r\leq k$ of distinct indices used and uses the bound 
$\binom{N}{r}\leq(eN/r)^{r}$. The regime $2k>N$ is handled by the trivial bound. 
Details are given in Appendix~\ref{app:proof_endpoint_counting}.

\subsection{Proof of Theorem~\ref{thm:quantum_entropy_poc}}
\label{subsec:proof_main_theorem}
Before stating the proof, we set
\begin{align*}
\mathbf{M}_t :=  m_t^{\otimes N},\quad 
\Lambda_{t} := \log(m_t).
\end{align*}
and we define operators, 
\begin{align}
\mathfrak{a}_{t}&:=-\mathrm{i}[A_{12},\Lambda_{t}\otimes\mathbf{1}+\mathbf{1}\otimes\Lambda_{t}],\label{eq:def_a_t}\\
\mathfrak{b}_{t}&:=\mathrm{tr}_{2}\bigl((\mathbf{1}\otimes m_{t})\mathfrak{a}_{t}\bigr),\label{eq:def_b_t}\\
\widehat{\mathfrak{a}}_{t}&:=\mathfrak{a}_{t}-\mathfrak{b}_{t}\otimes\mathbf{1}-\mathbf{1}\otimes\mathfrak{b}_{t}.\label{eq:def_a_hat_t}
\end{align}

The proof proceeds in three steps. We first compute the time derivative of $\mathfrak{H}_{N}(t)$ and use the monotonicity of the relative entropy under the CPTP semigroup $(e^{s\mathcal{L}^{N}})_{s\geq 0}$ to absorb the dissipative contributions. We then exploit the locality of $\mathbf{A}_{lr}$ to reduce the analysis to the centered two-body operator $\widehat{\mathfrak{a}}_{t}$. The third step controls the resulting expression by an exponential moment estimate via~\eqref{eq:variational_GT} and Lemmas~\ref{lem:isolated_vertex}--\ref{lem:endpoint_counting}.

We start taking the derivative of the normalize relative entropy,
by~\eqref{eq:def_T_sigma}--\eqref{eq:dt_log}, $\frac{\mathrm{d}}{\mathrm{d}t}\log \mathbf{M}_t=\mathcal{T}_{\mathbf{M}_t}(\dot{\mathbf{M}}_t)$, then 
\begin{align*}
\mathfrak{H}_{N}(t)=\frac{1}{N}\,\mathrm{tr}\bigl(\bar{\boldsymbol{\rho}}_t^{N}\log\bar{\boldsymbol{\rho}}_t^{N}-\bar{\boldsymbol{\rho}}_t^{N}\log \mathbf{M}_t\bigr).
\end{align*}
Differentiating the first term and using~\eqref{eq:dt_log} together with~\eqref{eq:trace_T}, we obtain
\begin{align*}
\frac{\mathrm{d}}{\mathrm{d}t}\mathrm{tr}(\bar{\boldsymbol{\rho}}_t^{N}\log\bar{\boldsymbol{\rho}}_t^{N})
&=\mathrm{tr}(\dot{\bar{\boldsymbol{\rho}}}_t^{N}\log\bar{\boldsymbol{\rho}}_t^{N})
+\mathrm{tr}\bigl(\bar{\boldsymbol{\rho}}_t^{N}\,
\mathcal{T}_{\bar{\boldsymbol{\rho}}_t^{N}}(\dot{\bar{\boldsymbol{\rho}}}_t^{N})\bigr)\\
&=\mathrm{tr}(\dot{\bar{\boldsymbol{\rho}}}_t^{N}\log\bar{\boldsymbol{\rho}}_t^{N})
+\mathrm{tr}\bigl(\mathcal{T}_{\bar{\boldsymbol{\rho}}_t^{N}}(\bar{\boldsymbol{\rho}}_t^{N})
\dot{\bar{\boldsymbol{\rho}}}_t^{N}\bigr)\\
&=\mathrm{tr}(\dot{\bar{\boldsymbol{\rho}}}_t^{N}\log\bar{\boldsymbol{\rho}}_t^{N})
+\mathrm{tr}(\dot{\bar{\boldsymbol{\rho}}}_t^{N})\\
&=\mathrm{tr}(\dot{\bar{\boldsymbol{\rho}}}_t^{N}\log\bar{\boldsymbol{\rho}}_t^{N}),
\end{align*}
where we used
$\mathcal{T}_{\bar{\boldsymbol{\rho}}_t^{N}}(\bar{\boldsymbol{\rho}}_t^{N})=\mathbf{1}$
and $\mathrm{tr}(\dot{\bar{\boldsymbol{\rho}}}_t^{N})=0$.
Differentiating the second term yields
\begin{align*}
\frac{\mathrm{d}}{\mathrm{d}t}\mathrm{tr}(\bar{\boldsymbol{\rho}}_t^{N}\log \mathbf{M}_t)=\mathrm{tr}(\dot{\bar{\boldsymbol{\rho}}}_t^{N}\log \mathbf{M}_t)+\mathrm{tr}\bigl(\bar{\boldsymbol{\rho}}_t^{N}\,\mathcal{T}_{\mathbf{M}_t}(\dot{\mathbf{M}}_t)\bigr).
\end{align*}
Combining and using $\dot{\bar{\boldsymbol{\rho}}}_t^{N}=\mathcal{L}^{N}(\bar{\boldsymbol{\rho}}_t^{N})$,
\begin{align}\label{eq:dt_HN_first}
\mathfrak{H}_{N}'(t)=\frac{1}{N}\Bigl[\mathrm{tr}\bigl(\mathcal{L}^{N}(\bar{\boldsymbol{\rho}}_t^{N})(\log\bar{\boldsymbol{\rho}}_t^{N}-\log \mathbf{M}_t)\bigr)-\mathrm{tr}\bigl(\bar{\boldsymbol{\rho}}_t^{N}\,\mathcal{T}_{\mathbf{M}_t}(\dot{\mathbf{M}}_t)\bigr)\Bigr].
\end{align}

We compare the two terms in~\eqref{eq:dt_HN_first} via the monotonicity of the relative entropy under $(e^{s\mathcal{L}^{N}})_{s\geq 0}$. Fix $t\in[0,T]$ and define
\begin{align*}
g(s):=\mathfrak{D}\bigl(e^{s\mathcal{L}^{N}}[\bar{\boldsymbol{\rho}}_t^{N}]\,\big\|\,e^{s\mathcal{L}^{N}}[\mathbf{M}_t]\bigr),\qquad s\geq 0.
\end{align*}
Since $(e^{s\mathcal{L}^{N}})_{s\geq 0}$ is a CPTP semigroup, $g(s)\leq g(0)$, hence $g'(0^{+})\leq 0$. Setting $\rho_{s}^{t}:=e^{s\mathcal{L}^{N}}[\bar{\boldsymbol{\rho}}_t^{N}] $ and $M_{s}^{t}:=e^{s\mathcal{L}^{N}}[\mathbf{M}_t]$, the same differentiation but in $s$ yields
\begin{align*}
\frac{\mathrm{d}}{\mathrm{d}s}\mathfrak{D}(\rho_{s}^{t}\|M_{s}^{t})=\mathrm{tr}\bigl(\mathcal{L}^{N}(\rho_{s}^{t})(\log\rho_{s}^{t}-\log M_{s}^{t})\bigr)-\mathrm{tr}\bigl(\rho_{s}^{t}\,\mathcal{T}_{M_{s}^{t}}(\mathcal{L}^{N}(M_{s}^{t}))\bigr).
\end{align*}
Evaluating at $s=0$ and using $g'(0^{+})\leq 0$, we obtain
\begin{align}\label{eq:monotonicity_consequence}
\mathrm{tr}\bigl(\mathcal{L}^{N}(\bar{\boldsymbol{\rho}}_t^{N})(\log\bar{\boldsymbol{\rho}}_t^{N}-\log \mathbf{M}_t)\bigr)\leq\mathrm{tr}\bigl(\bar{\boldsymbol{\rho}}_t^{N}\,\mathcal{T}_{\mathbf{M}_t}(\mathcal{L}^{N}(\mathbf{M}_t))\bigr).
\end{align}
Substituting~\eqref{eq:monotonicity_consequence} into~\eqref{eq:dt_HN_first},
\begin{align}\label{eq:HN_after_monotonicity}
\mathfrak{H}_{N}'(t)\leq\frac{1}{N}\,\mathrm{tr}\Bigl(\bar{\boldsymbol{\rho}}_t^{N}\,\mathcal{T}_{\mathbf{M}_t}\bigl(\mathcal{L}^{N}(\mathbf{M}_t)-\dot{\mathbf{M}}_t\bigr)\Bigr).
\end{align}

We now compute $\mathcal{L}^{N}(\mathbf{M}_t)-\dot{\mathbf{M}}_t$. By the product rule for tensor products,
\begin{align*}
\dot{\mathbf{M}}_t=\sum_{l=1}^{N}m_{t}^{\otimes(l-1)}\otimes\dot m_{t}\otimes m_{t}^{\otimes(N-l)}.
\end{align*}
Using $\dot m_{t}=-\mathrm{i}[\tilde H+A^{m_{t}},m_{t}]+\mathcal{D}(m_{t})$, we get
\begin{align*}
\dot{\mathbf{M}}_t=-\mathrm{i}\Bigl[\sum_{l=1}^{N}\mathbf{\tilde H}_{l}+\sum_{l=1}^{N}\mathbf{A}_{l}^{m_{t}},\,\mathbf{M}_t\Bigr]+\sum_{l=1}^{N}\mathcal{D}_{l}(\mathbf{M}_t).
\end{align*}
On the other hand, by~\eqref{eq:N_body_lindblad}, we have
\begin{align*}
\mathcal{L}^{N}(\mathbf{M}_t)=-\mathrm{i}[\mathbf{H}^{N},\mathbf{M}_t]+\sum_{l=1}^{N}\mathcal{D}_{l}(\mathbf{M}_t).
\end{align*}
The free Hamiltonian and dissipative parts then cancel, yielding
\begin{align}\label{eq:defect}
\mathcal{L}^{N}(\mathbf{M}_t)-\dot{\mathbf{M}}_t=-\mathrm{i}[\Delta_{t}^{N},\mathbf{M}_t],\qquad\Delta_{t}^{N}:=\frac{1}{N}\sum_{1\leq l<r\leq N}\mathbf{A}_{lr}-\sum_{l=1}^{N}\mathbf{A}_{l}^{m_{t}}.
\end{align}

Substituting~\eqref{eq:defect} into~\eqref{eq:HN_after_monotonicity} and applying~\eqref{eq:T_commutator} with $\sigma=\mathbf{M}_t$ and $X=\Delta_{t}^{N}$, we get
\begin{align*}
\mathcal{T}_{\mathbf{M}_t}\bigl(-\mathrm{i}[\Delta_{t}^{N},\mathbf{M}_t]\bigr)=-\mathrm{i}[\Delta_{t}^{N},\log \mathbf{M}_t].
\end{align*}
Hence
\begin{align}\label{eq:HN_commutator_form}
\mathfrak{H}_{N}'(t)\leq\frac{1}{N}\,\mathrm{tr}\Bigl(\bar{\boldsymbol{\rho}}_t^{N}(-\mathrm{i})[\Delta_{t}^{N},\log \mathbf{M}_t]\Bigr).
\end{align}
Since $\log \mathbf{M}_t=\sum_{k=1}^{N}\mathbf{\Lambda}_{t,k}$, and since $\mathbf{A}_{lr}$ commutes with $\mathbf{\Lambda}_{t,k}$ for $k\neq l,r$ while $\mathbf{A}_{l}^{m_{t}}$ commutes with $\mathbf{\Lambda}_{t,k}$ for $k\neq l$,
\begin{align*}
[\mathbf{A}_{lr},\log \mathbf{M}_t]=[\mathbf{A}_{lr},\mathbf{\Lambda}_{t,l}+\mathbf{\Lambda}_{t,r}],\qquad[\mathbf{A}_{l}^{m_{t}},\log \mathbf{M}_t]=[\mathbf{A}_{l}^{m_{t}},\mathbf{\Lambda}_{t,l}].
\end{align*}
Inserting these into~\eqref{eq:HN_commutator_form} and exploiting exchangeability, each pairwise term contributes $\mathrm{tr}(\bar{\boldsymbol{\rho}}_t^{N:(2)}\mathfrak{a}_{t})$ and each one-body term contributes $\mathrm{tr}\bigl(\bar{\boldsymbol{\rho}}_t^{N:(1)}(-\mathrm{i})[A^{m_{t}},\Lambda_{t}]\bigr)$. Counting yields
\begin{align}\label{eq:HN_two_terms}
\mathfrak{H}_{N}'(t)\leq\frac{N-1}{2N}\,\mathrm{tr}\bigl(\bar{\boldsymbol{\rho}}_t^{N:(2)}\mathfrak{a}_{t}\bigr)-\mathrm{tr}\Bigl(\bar{\boldsymbol{\rho}}_t^{N:(1)}(-\mathrm{i})[A^{m_{t}},\Lambda_{t}]\Bigr).
\end{align}
The second term equals $\mathrm{tr}(\bar{\boldsymbol{\rho}}_t^{N:(1)}\mathfrak{b}_{t})$. This can be shown by remarking that 
\begin{align}\label{eq:b_t_identification}
\mathfrak{b}_{t}=-\mathrm{i}[A^{m_{t}},\Lambda_{t}].
\end{align} 
See the details of the calculations in Appendix~\ref{app:proof_b_t_identification}. Hence
\begin{align*}
\mathfrak{H}_{N}'(t)\leq\frac{N-1}{2N}\,\mathrm{tr}\bigl(\bar{\boldsymbol{\rho}}_t^{N:(2)}\mathfrak{a}_{t}\bigr)-\mathrm{tr}\bigl(\bar{\boldsymbol{\rho}}_t^{N:(1)}\mathfrak{b}_{t}\bigr).
\end{align*}
By decomposing $\mathfrak{a}_{t}=\widehat{\mathfrak{a}}_{t}+\mathfrak{b}_{t}\otimes\mathbf{1}+\mathbf{1}\otimes\mathfrak{b}_{t}$ as in~\eqref{eq:def_a_hat_t}, and using $\mathrm{tr}(\bar{\boldsymbol{\rho}}_t^{N:(2)}(\mathfrak{b}_{t}\otimes\mathbf{1}))=\mathrm{tr}(\bar{\boldsymbol{\rho}}_t^{N:(2)}(\mathbf{1}\otimes\mathfrak{b}_{t}))=\mathrm{tr}(\bar{\boldsymbol{\rho}}_t^{N:(1)}\mathfrak{b}_{t})$, we get
\begin{align}\label{eq:HN_centered_form}
\mathfrak{H}_{N}'(t)\leq\frac{N-1}{2N}\,\mathrm{tr}\bigl(\bar{\boldsymbol{\rho}}_t^{N:(2)}\widehat{\mathfrak{a}}_{t}\bigr)+\frac{\|\mathfrak{b}_{t}\|}{N}.
\end{align}
Now  by~\eqref{eq:b_t_identification} and $\|A^{m_{t}}\|\leq\|A\|,$ we have   
\begin{align*}
\|\mathfrak{b}_{t}\|\leq 2\|A^{m_{t}}\|\,\|\Lambda_{t}\|\leq 2\|A\|\mathrm{K}.
\end{align*}

Now we try to control the exponential moment estimate.
We introduce
\begin{align*}
\mathfrak{U}_{N,t}:=\frac{1}{N^{2}}\sum_{1\leq i<j\leq N}(\widehat{\mathfrak{a}}_{t})_{ij},\qquad\mathfrak{W}_{N,t}:=N\mathfrak{U}_{N,t}=\frac{1}{N}\sum_{1\leq i<j\leq N}(\widehat{\mathfrak{a}}_{t})_{ij}.
\end{align*}
By exchangeability of $\bar{\boldsymbol{\rho}}_t^{N}$,
\begin{align*}
\mathrm{tr}(\bar{\boldsymbol{\rho}}_t^{N}\mathfrak{U}_{N,t})=\frac{N-1}{2N}\,\mathrm{tr}\bigl(\bar{\boldsymbol{\rho}}_t^{N:(2)}\widehat{\mathfrak{a}}_{t}\bigr),
\end{align*}
hence~\eqref{eq:HN_centered_form} reads
\begin{align}\label{eq:HN_with_U}
\mathfrak{H}_{N}'(t)\leq\mathrm{tr}(\bar{\boldsymbol{\rho}}_t^{N}\mathfrak{U}_{N,t})+\frac{\|\mathfrak{b_t}\|}{N}.
\end{align}
Applying~\eqref{eq:variational_GT} with $\rho=\bar{\boldsymbol{\rho}}_t^{N}$, $\sigma=\mathbf{M}_t$, $X=\mathfrak{U}_{N,t}$, $\lambda=\mathfrak{q}N$,
\begin{align}\label{eq:variational_applied}
\mathrm{tr}(\bar{\boldsymbol{\rho}}_t^{N}\mathfrak{U}_{N,t})\leq\frac{1}{\mathfrak{q}}\,\mathfrak{H}_{N}(t)+\frac{1}{\mathfrak{q}N}\log\mathrm{tr}\bigl(\mathbf{M}_t\,e^{\mathfrak{q}\mathfrak{W}_{N,t}}\bigr).
\end{align}

It remains to bound $\mathrm{tr}(\mathbf{M}_te^{\mathfrak{q}\mathfrak{W}_{N,t}})$ uniformly in $N$. For every $k\geq 1$, expanding $\mathfrak{W}_{N,t}^{k}=\bigl(\frac{1}{N}\sum_{1\leq i<j\leq N}(\widehat{\mathfrak{a}}_{t})_{ij}\bigr)^{k}$ as a sum over edges $\mathrm{e}_{1},\dots,\mathrm{e}_{k}$, where each $\mathrm{e}_{r}=\{i_{r},j_{r}\}$ with $1\leq i_{r}<j_{r}\leq N$, gives
\begin{align*}
\mathrm{tr}\bigl(\mathbf{M}_{t}\,\mathfrak{W}_{N,t}^{k}\bigr)=\frac{1}{N^{k}}\sum_{\mathrm{e}_{1},\dots,\mathrm{e}_{k}}\mathrm{tr}\bigl(m_{t}^{\otimes N}\,\widehat{\mathfrak{a}}_{t,\mathrm{e}_{1}}\cdots\widehat{\mathfrak{a}}_{t,\mathrm{e}_{k}}\bigr),
\end{align*}
where the sum runs over all $k$-tuples of edges in $\{1,\dots,N\}$, and $\widehat{\mathfrak{a}}_{t,\mathrm{e}_{r}}=(\widehat{\mathfrak{a}}_{t})_{i_{r}j_{r}}$ denotes the copy of $\widehat{\mathfrak{a}}_{t}$ acting on the $i_{r}$-th and $j_{r}$-th tensor factors of $\mathbb{H}^{\otimes N}$.

We first observe that  $\widehat{\mathfrak{a}}_t$ satisfies the assumption of Lemma~\ref{lem:isolated_vertex}  with $\varrho=m_{t}$, (see details in Appendix~\ref{app:proof_centering}), so the lemma can be applied, and this eliminates every term in which one index appears exactly once. The surviving ordered $k$-tuples are in $\mathfrak{K}_{N,2k} := \Bigl\{
(i_{1},j_{1},\dots,i_{k},j_{k})\in\{1,\dots,N\}^{2k}
:\text{no index appears exactly once}
\Bigr\} $, and
\begin{align}\label{eq:moment_bound_K}
\bigl|\mathrm{tr}\bigl(\mathbf{M}_t\mathfrak{W}_{N,t}^{k}\bigr)\bigr|\leq\frac{|\mathfrak{K}_{N,2k}|}{N^{k}}\|\widehat{\mathfrak{a}}_t\|^{k}.
\end{align}

We now distinguish two regimes.

If $2k\leq N$, Lemma~\ref{lem:endpoint_counting} gives $|\mathfrak{K}_{N,2k}|\leq k\,e^{k}\,N^{k}\,k^{k}$, hence
\begin{align*}
\bigl|\mathrm{tr}\bigl(\mathbf{M}_t\mathfrak{W}_{N,t}^{k}\bigr)\bigr|\leq k\,e^{k}\,k^{k}\,\|\widehat{\mathfrak{a}}_{t}\|^{k},
\end{align*}
and using $k!\geq(k/e)^{k}$,
\begin{align}\label{eq:moment_bound_small_k}
\frac{\mathfrak{q}^{k}}{k!}\bigl|\mathrm{tr}\bigl(\mathbf{M}_t\mathfrak{W}_{N,t}^{k}\bigr)\bigr|\leq k\bigl(e^{2}\mathfrak{q}\,\|\widehat{\mathfrak{a}}_{t}\|\bigr)^{k}.
\end{align}
If $2k>N$, then $|\mathfrak{K}_{N,2k}|\leq N^{2k}$ and $N<2k$, hence
\begin{align*}
\bigl|\mathrm{tr}\bigl(\mathbf{M}_t\mathfrak{W}_{N,t}^{k}\bigr)\bigr|\leq N^{k}\|\widehat{\mathfrak{a}}_t\|^{k}\leq(2k)^{k}\|\widehat{\mathfrak{a}}_{t}\|^{k},
\end{align*}
then using Stirling's, 
\begin{align}\label{eq:moment_bound_large_k}
\frac{\mathfrak{q}^{k}}{k!}\bigl|\mathrm{tr}\bigl(\mathbf{M}_t\mathfrak{W}_{N,t}^{k}\bigr)\bigr|\leq\bigl(2e\mathfrak{q}\,\|\widehat{\mathfrak{a}}_{t}\|\bigr)^{k}.
\end{align}

We now sum the exponential series:
\begin{align*}
\mathrm{tr}\bigl(
\boldsymbol{m}_t^N e^{\mathfrak{q}\mathfrak{W}_{N,t}}
\bigr)
&=
\sum_{k=0}^{\infty}
\frac{\mathfrak{q}^k}{k!}
\mathrm{tr}\bigl(\boldsymbol{m}_t^N\mathfrak{W}_{N,t}^k\bigr)
\\
&\leq
1
+
\sum_{\substack{k\geq 1\\2k\leq N}}
\frac{|\mathfrak{q}|^k}{k!}
\bigl|
\mathrm{tr}\bigl(\boldsymbol{m}_t^N\mathfrak{W}_{N,t}^k\bigr)
\bigr|
+
\sum_{\substack{k\geq 1\\2k>N}}
\frac{|\mathfrak{q}|^k}{k!}
\bigl|
\mathrm{tr}\bigl(\boldsymbol{m}_t^N\mathfrak{W}_{N,t}^k\bigr)
\bigr|,\\
&\leq 1 +
\sum_{\substack{k\geq 1\\2k\leq N}}
k\bigl(e^{2}\mathfrak{q}\,\|\widehat{\mathfrak{a}}_{t}\|\bigr)^{k}
+
\sum_{\substack{k\geq 1\\2k>N}}
\bigl(2e\mathfrak{q}\,\|\widehat{\mathfrak{a}}_{t}\|\bigr)^{k},\\
&\leq 1 +
\sum_{\substack{k\geq 1}}
k\bigl(e^{2}\mathfrak{q}\,\|\widehat{\mathfrak{a}}_{t}\|\bigr)^{k}
+
\sum_{\substack{k\geq 1}}
\bigl(2e\mathfrak{q}\,\|\widehat{\mathfrak{a}}_{t}\|\bigr)^{k}, 
\end{align*}

Using the fact that, $\|\mathfrak{a}_{t}\|\leq 2\|A\|\,\|\Lambda_{t}\otimes\mathbf{1}+\mathbf{1}\otimes\Lambda_{t}\|\leq 4\|A\|\mathrm{K}$, and~\eqref{eq:def_a_hat_t} gives
\begin{align*}
\|\widehat{\mathfrak{a}}_{t}\|\leq\|\mathfrak{a}_{t}\|+2\|\mathfrak{b}_{t}\|\leq 8\|A\|\mathrm{K}.
\end{align*}

Setting $x_{T,\mathfrak{q}}:=8e^{2}\mathfrak{q}\|A\|\mathrm{K}$ and $y_{T,\mathfrak{q}}:=16e\mathfrak{q}\|A\|\mathrm{K}$,
\begin{align*}
\mathrm{tr}\bigl(\mathbf{M}_t\,e^{\mathfrak{q}\mathfrak{W}_{N,t}}\bigr)\leq 1+\sum_{k\geq 1}k\,x_{T,\mathfrak{q}}^{k}+\sum_{k\geq 1}y_{T,\mathfrak{q}}^{k}.
\end{align*}
Since $\mathfrak{q}<\tfrac{e^{-2}}{8\|A\|\mathrm{K}}$, $x_{T,\mathfrak{q}}<1$, and since $y_{T,\mathfrak{q}}=(2/e)\,x_{T,\mathfrak{q}}<x_{T,\mathfrak{q}}<1$, the two series converges and there equal to : 
\begin{align*}
\sum_{k\geq 1}k\,x_{T,\mathfrak{q}}^{k}=\frac{x_{T,\mathfrak{q}}}{(1-x_{T,\mathfrak{q}})^{2}},\qquad\sum_{k\geq 1}y_{T,\mathfrak{q}}^{k}=\frac{y_{T,\mathfrak{q}}}{1-y_{T,\mathfrak{q}}}.
\end{align*}
Therefore
\begin{align}\label{eq:exp_moment_bound}
\mathrm{tr}\bigl(\mathbf{M}_t\,e^{\mathfrak{q}\mathfrak{W}_{N,t}}\bigr)\leq C_{T,\mathfrak{q}},
\end{align}
with $C_{T,\mathfrak{q}}$ as in~\eqref{eq:def_C_T_q}.
Combining~\eqref{eq:HN_with_U}, \eqref{eq:variational_applied}, and~\eqref{eq:exp_moment_bound},
\begin{align*}
\mathfrak{H}_{N}'(t)\leq\frac{1}{\mathfrak{q}}\,\mathfrak{H}_{N}(t)+\frac{\log C_{T,\mathfrak{q}}}{\mathfrak{q}N}+\frac{2\|A\|\mathrm{K}}{N}.
\end{align*}
Grönwall's lemma yields
\begin{align*}
\mathfrak{H}_{N}(t)\leq e^{t/\mathfrak{q}}\left(\mathfrak{H}_{N}(0)+\frac{\log C_{T,\mathfrak{q}}+2\mathfrak{q}\|A\|\mathrm{K}}{N}\right),
\end{align*}
which is~\eqref{eq:main_estimate}.
\hfill \qedsymbol
\subsection{Proof of Corollary~\ref{cor:entropy_to_marginals}}
\label{subsec:proof_corollary} 
Fix $k\geq 1$, and assume without loss of generality that $N\geq 2k$. Set $n:=\lfloor N/k\rfloor$, so that $nk\leq N < (n+1)k$.

By the monotonicity of the relative entropy under partial trace~\eqref{eq:monotonicity_partial_trace}, applied iteratively to trace out the last $N-nk$ factors, we obtain
\begin{align}\label{eq:cor_step1}
\mathfrak{D}\bigl(\bar{\boldsymbol\rho}_{t}^{N}\,\big\|\,m_{t}^{\otimes N}\bigr)
\geq
\mathfrak{D}\bigl(\bar{\boldsymbol\rho}_{t}^{N:(nk)}\,\big\|\,m_{t}^{\otimes nk}\bigr).
\end{align}
We next decompose $\{1,\dots,nk\}$ into $n$ disjoint consecutive blocks
\begin{align*}
B_{j} := \{(j-1)k+1,\dots,jk\}, \qquad j=1,\dots,n,
\end{align*}
each of size $k$. Applying the superadditivity property~\eqref{eq:superadditivity} to this decomposition yields
\begin{align*}
\mathfrak{D}\bigl(\bar{\boldsymbol\rho}_{t}^{N:(nk)}\,\big\|\,m_{t}^{\otimes nk}\bigr)
\geq
\sum_{j=1}^{n}\mathfrak{D}\bigl(\bar{\boldsymbol\rho}_{t}^{N:(k),B_{j}}\,\big\|\,m_{t}^{\otimes k}\bigr).
\end{align*}
By exchangeability, we have
\begin{align*}
\bar{\boldsymbol\rho}_{t}^{N:(k),B_{j}} = \bar{\boldsymbol\rho}_{t}^{N:(k)},\qquad j=1,\dots,n.
\end{align*}
Substituting this into the previous inequality, we obtain
\begin{align}\label{eq:cor_step2}
\mathfrak{D}\bigl(\bar{\boldsymbol\rho}_{t}^{N:(nk)}\,\big\|\,m_{t}^{\otimes nk}\bigr)
\geq
n\,\mathfrak{D}\bigl(\bar{\boldsymbol\rho}_{t}^{N:(k)}\,\big\|\,m_{t}^{\otimes k}\bigr).
\end{align}
Combining~\eqref{eq:cor_step1} and~\eqref{eq:cor_step2}, and recalling the definition $\mathfrak{H}_N(t) = \frac{1}{N}\mathfrak{D}(\bar{\boldsymbol\rho}_t^N\|m_t^{\otimes N})$ given in~\eqref{eq:def_HN}, we obtain
\begin{align*}
n\,\mathfrak{D}\bigl(\bar{\boldsymbol\rho}_{t}^{N:(k)}\,\big\|\,m_{t}^{\otimes k}\bigr)
\leq
\mathfrak{D}\bigl(\bar{\boldsymbol\rho}_{t}^{N}\,\big\|\,m_{t}^{\otimes N}\bigr)
= N\,\mathfrak{H}_{N}(t).
\end{align*}
Dividing by $n=\lfloor N/k\rfloor$ and using $\lfloor N/k\rfloor\geq N/(2k)$ (which holds since $N\geq 2k$), we get
\begin{align}\label{eq:cor_bound}
\mathfrak{D}\bigl(\bar{\boldsymbol\rho}_{t}^{N:(k)}\,\big\|\,m_{t}^{\otimes k}\bigr)
\leq
\frac{N}{\lfloor N/k\rfloor}\,\mathfrak{H}_{N}(t)
\leq 2k\,\mathfrak{H}_{N}(t).
\end{align}
By Theorem~\ref{thm:quantum_entropy_poc}, for any fixed $0<\mathfrak{q}<\frac{e^{-2}}{8\|A\|\mathrm{K}}$, we have,  for every $t\in[0,T]$,
\begin{align*}
\mathfrak{H}_{N}(t)
\leq e^{T/\mathfrak{q}}\!\left(\mathfrak{H}_{N}(0)+\frac{\log C_{T,\mathfrak{q}}+2\mathfrak{q}\|A\|\mathrm{K}}{N}\right).
\end{align*}
The right-hand side is independent of $t$ and tends to $0$ as $N\to\infty$ under the assumption that $\mathfrak{H}_{N}(0)\to 0$. Hence
\begin{align*}
\sup_{t\in[0,T]}\mathfrak{H}_{N}(t)\xrightarrow[N\to\infty]{}0,
\end{align*}
and taking the supremum in~\eqref{eq:cor_bound} yields~\eqref{eq:cor_entropy_marginal}.

Finally, the quantum Pinsker inequality~\eqref{eq:pinsker}, applied to $\bar{\boldsymbol\rho}_{t}^{N:(k)}$ and $m_{t}^{\otimes k}$, gives
\begin{align*}
\frac{1}{2}\,\bigl\|\bar{\boldsymbol\rho}_{t}^{N:(k)}-m_{t}^{\otimes k}\bigr\|_{1}^{2}
\leq
\mathfrak{D}\bigl(\bar{\boldsymbol\rho}_{t}^{N:(k)}\,\big\|\,m_{t}^{\otimes k}\bigr).
\end{align*}
Combining this with~\eqref{eq:cor_bound}, we obtain
\begin{align*}
\sup_{t\in[0,T]}\bigl\|\bar{\boldsymbol\rho}_{t}^{N:(k)}-m_{t}^{\otimes k}\bigr\|_{1}^{2}
\leq
4k\,\sup_{t\in[0,T]}\mathfrak{H}_{N}(t)
\xrightarrow[N\to\infty]{}0,
\end{align*}
which proves~\eqref{eq:cor_trace_norm}.
\hfill \qedsymbol
\section{Conclusion}
\label{sec:conclusion}

The use of the Umegaki relative entropy has allowed us to extend propagation of chaos to open quantum systems. Convergence of the entropy yields, in an essentially uniform way, convergence of all marginals together with an explicit rate of convergence. A specific feature of the method is that it applies naturally to symmetric mixed states, which makes it particularly well suited to open quantum dynamics.

Future work will consider the case involving quantum measurements, where stochastic terms are added to the Lindblad equation. Furthermore, we will study the infinite-dimensional case, where we believe the relative entropy estimate can also be adapted. 

More precisely, in finite dimension, the faithfulness of $m_{t}$ ensures that $\log m_{t}$ is a bounded operator, so that the constant $\sup_{t\in[0,T]}\|\log(m_{t})\| < \infty$. In infinite dimension, this is no longer true, their eigenvalues accumulate at zero and $\log m_{t}$ is generically unbounded as an operator. The bound can be replaced by a finite-moment condition that allows one to control the exponential moments appearing in~\eqref{eq:variational_applied}.

A natural such condition is, for some $\theta \in (0,1)$,
\begin{align}\label{eq:exponential_moment_condition}
\sup_{t\in[0,T]}\mathrm{tr}\bigl(e^{\theta|\Lambda_{t}|}m_{t}\bigr)<+\infty.
\end{align}
This condition is satisfied, for instance, by the Gibbs state on a single bosonic mode, which is a fundamental example in quantum statistical mechanics and many-body physics~\cite{breuer_petruccione,bratteli12operator}.
 With $\mathbb{H}=\ell^{2}(\mathbb{N})$ and number operator $\mathsf{N}$,
\begin{align*}
\varrho_{\beta}
=
\frac{e^{-\beta\mathsf{N}}}{\mathrm{tr}(e^{-\beta\mathsf{N}})}
=
(1-e^{-\beta})\sum_{n\geq 0}e^{-\beta n}|n\rangle\langle n|,
\qquad\beta>0,
\end{align*}
yields
\begin{align*}
\log\varrho_{\beta}=\log(1-e^{-\beta})\mathbf{1}-\beta\mathsf{N},
\end{align*}
which is unbounded. However, the eigenvalues are $p_{n}=(1-e^{-\beta})e^{-\beta n}$, and hence for every $\theta\in(0,1)$,
\begin{align*}
\mathrm{tr}\bigl(e^{\theta|\log\varrho_{\beta}|}\varrho_{\beta}\bigr)
=
\sum_{n\geq 0}p_{n}^{\,1-\theta}
=
(1-e^{-\beta})^{1-\theta}\sum_{n\geq 0}e^{-\beta(1-\theta)n}
<+\infty.
\end{align*}
This example suggests that exponential moment bounds of the form~\eqref{eq:exponential_moment_condition} provide a natural replacement for the boundedness of $\log m_t$ in infinite dimension. Establishing propagation of chaos under such conditions will be the object of future work.

{\bf Acknowledgments:} This work was supported by ANR-19-CE48-0003 and ANR-21-CE47-0015. The authors thank the Institute for Mathematical and Statistical Innovation (IMSI), supported by the National Science Foundation (Grant No. DMS-1929348), and the International Centre for Theoretical Sciences (ICTS) \emph{“Quantum Trajectories”} program, where part of this research was conducted.
\bibliographystyle{plain} 

\bibliography{Refs}

\appendix
\section*{Appendices}
\addcontentsline{toc}{section}{Appendices}

\section{Proof of Proposition~\ref{prop:faithfulness}}
\label{app:proof_faithfulness}

Set $H_{t}:=\tilde{H}+A^{m_{t}}$ and
$K_{t}:=H_{t}-\frac{\mathrm{i}}{2}L^{\dagger}L\in\mathcal{B}(\mathbb{H}),$
so that
\begin{align}\label{eq:K_minus_Kdagger}
K_{t}-K_{t}^{\dagger}=-\mathrm{i}\,L^{\dagger}L.
\end{align}
A direct expansion using~\eqref{eq:K_minus_Kdagger} rewrites~\eqref{eq:mf_lindblad} as
\begin{align}\label{eq:mf_with_K}
\frac{\mathrm{d}m_{t}}{\mathrm{d}t}
=
-\mathrm{i}\bigl(K_{t}m_{t}-m_{t}K_{t}^{\dagger}\bigr)+L\,m_{t}\,L^{\dagger}.
\end{align}
Since $t\mapsto m_{t}$ is continuous on $[0,T]$, $t\mapsto K_{t}$ is continuous, hence
the linear ODE
\begin{align*}
\frac{\mathrm{d}V_{t}}{\mathrm{d}t}=-\mathrm{i}\,K_{t}\,V_{t},\qquad V_{0}=\mathbf{1},
\end{align*}
admits a unique global solution $V_{t}\in\mathcal{B}(\mathbb{H})$ on $[0,T]$. By
Liouville's formula,
\begin{align*}
\det V_{t}=\exp\!\int_{0}^{t}\mathrm{tr}\bigl(-\mathrm{i}K_{r}\bigr)\,\mathrm{d}r\neq 0,
\end{align*}
so $V_{t}$ is invertible for every $t\in[0,T]$.

Setting $W_{t,s}:=V_{t}\,V_{s}^{-1}$, one has $W_{s,s}=\mathbf{1}$ and
$\partial_{t}W_{t,s}=-\mathrm{i}K_{t}\,W_{t,s}$. Define
\begin{align*}
F_{t}:=V_{t}^{-1}\,m_{t}\,(V_{t}^{\dagger})^{-1}.
\end{align*}
Using $\partial_{t}V_{t}^{-1}=-V_{t}^{-1}(-\mathrm{i}K_{t})=\mathrm{i}V_{t}^{-1}K_{t}$
together with~\eqref{eq:mf_with_K},
\begin{align*}
\dot F_{t}
&=
\mathrm{i}V_{t}^{-1}K_{t}\,m_{t}\,(V_{t}^{\dagger})^{-1}
+
V_{t}^{-1}\dot m_{t}\,(V_{t}^{\dagger})^{-1}
-
\mathrm{i}V_{t}^{-1}\,m_{t}\,K_{t}^{\dagger}(V_{t}^{\dagger})^{-1}
\\
&=
V_{t}^{-1}\bigl(\dot m_{t}+\mathrm{i}K_{t}m_{t}-\mathrm{i}m_{t}K_{t}^{\dagger}\bigr)(V_{t}^{\dagger})^{-1}
=
V_{t}^{-1}\,L\,m_{t}\,L^{\dagger}\,(V_{t}^{\dagger})^{-1}.
\end{align*}
Integrating and conjugating by $V_{t}$ on the left and $V_{t}^{\dagger}$ on the right, we obtain
\begin{align}\label{eq:duhamel_proof}
m_{t}
=
V_{t}\,m_{0}\,V_{t}^{\dagger}
+
\int_{0}^{t}W_{t,s}\,L\,m_{s}\,L^{\dagger}\,W_{t,s}^{\dagger}\,\mathrm{d}s.
\end{align}
For every $s\in[0,t]$, $L\,m_{s}\,L^{\dagger}\geq 0$ since $m_{s}\geq 0$, and conjugation
by $W_{t,s}$ preserves positivity. The integrand in~\eqref{eq:duhamel_proof} is therefore
a positive operator, and so is the integral. Hence
\begin{align}\label{eq:m_t_geq_propagated}
m_{t}\geq V_{t}\,m_{0}\,V_{t}^{\dagger}.
\end{align}

Let $\psi\in\mathbb{H}$. Using~\eqref{eq:K_minus_Kdagger},
\begin{align*}
\frac{\mathrm{d}}{\mathrm{d}t}\|V_{t}\psi\|^{2}
&=
\bigl\langle V_{t}\psi,\,(-\mathrm{i}K_{t}+\mathrm{i}K_{t}^{\dagger})V_{t}\psi\bigr\rangle
=
-\bigl\langle V_{t}\psi,\,L^{\dagger}L\,V_{t}\psi\bigr\rangle
\\
&=
-\|L\,V_{t}\psi\|^{2}\geq -\|L\|^{2}\,\|V_{t}\psi\|^{2},
\end{align*}
where the anti-Hermitian part $-\mathrm{i}H_{t}$ does not contribute.
Gr\"onwall's lemma then yields, for every $\psi\in\mathbb{H}$,
\begin{align*}
\|V_{t}\psi\|^{2}\geq e^{-\|L\|^{2}t}\,\|\psi\|^{2},
\end{align*}
that is,
\begin{align*}
V_{t}^{\dagger}V_{t}\geq e^{-\|L\|^{2}t}\,\mathbf{1}.
\end{align*}
Since $V_{t}V_{t}^{\dagger}$ and $V_{t}^{\dagger}V_{t}$ have the same spectrum,
\begin{align}\label{eq:Vt_Vt_dagger_lower}
V_{t}\,V_{t}^{\dagger}\geq e^{-\|L\|^{2}t}\,\mathbf{1}.
\end{align}

Combining $m_{0}\geq\lambda_{\min}(m_{0})\,\mathbf{1}$
with~\eqref{eq:m_t_geq_propagated} and~\eqref{eq:Vt_Vt_dagger_lower}, we obtain
\begin{align*}
m_{t}
\geq
V_{t}\,m_{0}\,V_{t}^{\dagger}
\geq
\lambda_{\min}(m_{0})\,V_{t}V_{t}^{\dagger}
\geq
\lambda_{\min}(m_{0})\,e^{-\|L\|^{2}t}\,\mathbf{1}.
\end{align*}
Since $\lambda_{\min}(m_{0})>0$ by assumption, it follows that $m_{t}>0$ for every $t\in[0,T]$, i.e.\ $m_{t}$ is faithful and~\eqref{eq:lower_bound_eigenvalue} holds.

The eigenvalues of $m_{t}$ lie in $(0,1]$, so the spectrum of $\log m_{t}$ is contained
in $(-\infty,0],$ and therefore
\begin{align*}
\|\log m_{t}\|=-\log\lambda_{\min}(m_{t}).
\end{align*}
Combining this with~\eqref{eq:lower_bound_eigenvalue}, we obtain 
\begin{align*}
\|\log m_{t}\|\leq -\log\lambda_{\min}(m_{0})+\|L\|^{2}\,t.
\end{align*}
Taking the supremum over $t\in[0,T]$ proves~\eqref{eq:bound_Lambda_T}.
\hfill \qedsymbol

\section{Identification of $\mathfrak{b}_{t}$}
\label{app:proof_b_t_identification}

By the definition of $\mathfrak{b}_{t}$ given in Equation~\eqref{eq:def_b_t}, we have
\begin{align*}
\mathfrak{b}_{t}=-\mathrm{i}\,\mathrm{tr}_{2}\bigl((\mathbf{1}\otimes m_{t})[A_{12},\Lambda_{t}\otimes\mathbf{1}]\bigr)-\mathrm{i}\,\mathrm{tr}_{2}\bigl((\mathbf{1}\otimes m_{t})[A_{12},\mathbf{1}\otimes\Lambda_{t}]\bigr).
\end{align*}
Using the elementary partial-trace identities $\mathrm{tr}_{2}(X(B\otimes\mathbf{1}))=\mathrm{tr}_{2}(X)B$ and $\mathrm{tr}_{2}((B\otimes\mathbf{1})X)=B\,\mathrm{tr}_{2}(X)$, together with~\eqref{eq:def_A_sigma}, we obtain
\begin{align*}
\mathrm{tr}_{2}\bigl((\mathbf{1}\otimes m_{t})[A_{12},\Lambda_{t}\otimes\mathbf{1}]\bigr)=A^{m_{t}}\Lambda_{t}-\Lambda_{t}A^{m_{t}}=[A^{m_{t}},\Lambda_{t}].
\end{align*}
For the second term, since $m_{t}$ commutes with $\Lambda_{t}=\log m_{t}$, $(\mathbf{1}\otimes m_{t})$ commutes with $(\mathbf{1}\otimes\Lambda_{t})$, hence
\begin{align*}
\mathrm{tr}_{2}\bigl((\mathbf{1}\otimes m_{t})[A_{12},\mathbf{1}\otimes\Lambda_{t}]\bigr)=0.
\end{align*}
Combining the two identities yields $\mathfrak{b}_{t}=-\mathrm{i}[A^{m_{t}},\Lambda_{t}]$.

\section{Centering property of $\widehat{\mathfrak{a}}_{t}$}
\label{app:proof_centering}

We prove the two centering identities
\begin{align*}
\mathrm{tr}_{2}\bigl((\mathbf{1}\otimes m_{t})\widehat{\mathfrak{a}}_{t}\bigr)=0,
\qquad
\mathrm{tr}_{1}\bigl((m_{t}\otimes\mathbf{1})\widehat{\mathfrak{a}}_{t}\bigr)=0.
\end{align*}
By definition~\eqref{eq:def_a_hat_t} of $\widehat{\mathfrak{a}}_{t}$ and linearity of the partial trace,
\begin{align*}
\mathrm{tr}_{2}\bigl((\mathbf{1}\otimes m_{t})\widehat{\mathfrak{a}}_{t}\bigr)
=
\mathrm{tr}_{2}\bigl((\mathbf{1}\otimes m_{t})\mathfrak{a}_{t}\bigr)
-
\mathrm{tr}_{2}\bigl((\mathbf{1}\otimes m_{t})(\mathfrak{b}_{t}\otimes\mathbf{1})\bigr)
-
\mathrm{tr}_{2}\bigl((\mathbf{1}\otimes m_{t})(\mathbf{1}\otimes\mathfrak{b}_{t})\bigr).
\end{align*}
The first term equals $\mathfrak{b}_{t}$ by definition~\eqref{eq:def_b_t}, and the second term also equals  $\mathfrak{b}_{t}$ thanks to the identification $\mathfrak{b}_{t} = -\mathrm{i}[A^{m_t},\Lambda_t]$  and the third term equals $\mathrm{tr}(m_{t}\mathfrak{b}_{t})\mathbf{1}$. Combining,
\begin{align*}
\mathrm{tr}_{2}\bigl((\mathbf{1}\otimes m_{t})\widehat{\mathfrak{a}}_{t}\bigr)
=
\mathfrak{b}_{t}-\mathfrak{b}_{t}-\mathrm{tr}(m_{t}\mathfrak{b}_{t})\mathbf{1}
=
-\mathrm{tr}(m_{t}\mathfrak{b}_{t})\mathbf{1}.
\end{align*}
Then, 
\begin{align*}
\mathrm{tr}(m_{t}\mathfrak{b}_{t})
=
-\mathrm{i}\,\mathrm{tr}\bigl(m_{t}[A^{m_{t}},\Lambda_{t}]\bigr)
=
-\mathrm{i}\,\mathrm{tr}\bigl(m_{t}A^{m_{t}}\Lambda_{t}-m_{t}\Lambda_{t}A^{m_{t}}\bigr)
=
-\mathrm{i}\,\mathrm{tr}\bigl([\Lambda_{t},m_{t}]A^{m_{t}}\bigr)
=
0,
\end{align*}
where we used the cyclicity of the trace and the commutation $[\Lambda_{t},m_{t}]=[\log m_{t},m_{t}]=0$. This proves the first identity, and the proof is identical for $\mathrm{tr}_{1}\bigl((m_{t}\otimes\mathbf{1})\widehat{\mathfrak{a}}_{t}\bigr)$.

\section{Proof of Lemma~\ref{lem:isolated_vertex}}
\label{app:proof_isolated_vertex}

Assume that the index $l\in\{1,\dots,N\}$ appears exactly once among $i_{1},j_{1},\dots,i_{k},j_{k}$. Then, there exists a unique $s\in\{1,\dots,k\}$ such that $l\in \mathrm{e}_{s}$. Since $l$ appears nowhere else, every factor $h_{\mathrm{e}_{r}}$ with $r\neq s$ acts trivially on the $l$-th tensor factor. Hence we may write $h_{\mathrm{e}_{1}}\cdots h_{\mathrm{e}_{k}}=F\,h_{\mathrm{e}_{s}}\,G$, where $F$ and $G$ act trivially on the $l$-th factor. Identifying $\varrho^{\otimes N}=(\varrho)_{l}\otimes\varrho^{\otimes(N-1)}$ and tracing out the $l$-th component, we obtain
\begin{align*}
\mathrm{tr}\bigl(\varrho^{\otimes N}h_{\mathrm{e}_{1}}\cdots h_{\mathrm{e}_{k}}\bigr)=\mathrm{tr}\Bigl(\varrho^{\otimes(N-1)}F\,\mathrm{tr}_{l}\bigl((\varrho)_{l}h_{\mathrm{e}_{s}}\bigr)\,G\Bigr).
\end{align*}
If $l=i_{s}$, then $\mathrm{tr}_{l}((\varrho)_{l}h_{\mathrm{e}_{s}})$ is the transported copy of $\mathrm{tr}_{1}((\varrho\otimes\mathbf{1})h)=0$. If $l=j_{s}$, it is the corresponding copy of $\mathrm{tr}_{2}((\mathbf{1}\otimes\varrho)h)=0$. In both cases the inner partial trace vanishes, proving~\eqref{eq:isolated_vanishing}.

\hfill \qedsymbol

\section{Proof of Lemma~\ref{lem:endpoint_counting}}
\label{app:proof_endpoint_counting}

Take $\mathfrak{I}_{2k}:=(i_{1},j_{1},\dots,i_{k},j_{k})\in\mathfrak{K}_{N,2k}$ and let $r:=\|\mathfrak{I}_{2k}\|_{\mathrm{c}}$ denote the number of distinct indices appearing in $\mathfrak{I}_{2k}$. Since no index has multiplicity $1$, every distinct index appears at least twice, so $1\leq r\leq k$.

Decomposing $\mathfrak{K}_{N,2k}$ according to the value of $r$, choosing the $r$ values among $\{1,\dots,N\}$ in at most $\binom{N}{r}$ ways, then filling the $2k$ slots with these values in at most $r^{2k}$ ways, we obtain
\begin{align*}
|\mathfrak{K}_{N,2k}|\leq\sum_{r=1}^{k}\binom{N}{r}r^{2k}.
\end{align*}

If $2k\leq N$, then $k\leq N/2$, so the map $r\mapsto\binom{N}{r}$ is increasing on $\{1,\dots,k\}$ and $r^{2k}\leq k^{2k}$. Hence $|\mathfrak{K}_{N,2k}|\leq k\binom{N}{k}k^{2k}\leq k(eN/k)^{k}k^{2k}=k\,e^{k}\,N^{k}\,k^{k}$.

If $2k>N$, then trivially $|\mathfrak{K}_{N,2k}|\leq N^{2k}$.
\hfill \qedsymbol

\end{document}